\documentclass[preprint,prb]{revtex4}
\usepackage{amsmath}

\begin{document}

\title{Nodes of wavefunctions}

\author{M. Moriconi}

\affiliation{Departamento de F\'\i sica, Universidade Federal Fluminense, Av. Litor\^anea s/n, Boa Viagem - CEP 24210-340, Niter\'oi, Rio de Janeiro, Brazil}
\begin{abstract}
\noindent We give a simple argument to show that the $n$th wavefunction for the one-dimensional Schr\"odinger equation has $n-1$ nodes. We also show that if $n_1 < n_2$, then between two consecutive zeros of $\psi_{n_1}$, there is a zero of $\psi_{n_2}$.
\end{abstract}

\maketitle

The time-independent Schr{\"o}dinger equation in one dimension is a linear, second-order differential equation, which in units where $\hbar^2/2m=1$, can be expressed as
\begin{equation}
-\psi'' + V \psi = E \psi. \label{SE}
\end{equation}
All the properties of the wavefunction are encoded in Eq.~\eqref{SE}. We would like to know the properties of the zeros of its eigenfunctions, that is, how many nodes are there for the wavefunction $\psi_n$ corresponding to the $n$th energy level. Arguments based on the variational principle\cite{F,AJP} show that the ground state wavefunction has no nodes, but do not say much about wavefunctions for higher energy levels. This question is considered too detailed,\cite{L, GP} and is usually discussed in the more mathematical literature such as Refs.~\onlinecite{CH, H, M}. The number of zeros possessed by a solution of a linear, second-order, differential equation and their distribution is the subject of Sturm-Liouville theory.\cite{H} Two of the main theorems in Sturm-Liouville theory are the separation and the comparison theorems. For the Schr{\"o}dinger equation, the separation theorem states that the zeros of two linearly independent solutions of Eq.~(\ref{SE}) alternate; the comparison theorem states that if $n_1<n_2$, then between two consecutive zeros of $\psi_{n_1}$ there is a zero of $\psi_{n_2}$. There are stronger results that show that the solution corresponding to the $n$th eigenvalue has precisely $n-1$ zeros, and, if the potential goes to infinity as $|x| \to \infty$, then the eigenvalues form a discrete unbounded sequence. These theorems are very important in quantum mechanics, but they are usually not discussed in quantum mechanics textbooks.

The purpose of this note is to provide a simple, intuitive, argument for some of these results. We show that $\psi_n$ has $n-1$ nodes,\cite{note} and that if $n_2>n_1$, then between two zeros of $\psi_{n_1}$, there is a zero of $\psi_{n_2}$. The second result is simpler to prove, and we include it here for completeness. The first result is more difficult to prove and is the main contribution of this paper. The node structure of wavefunctions is important. For example, the fact that the ground state function has no nodes was one of the ingredients Feynman used in his theory of liquid Helium.\cite{F} In the following we assume that the normalizable solutions of Eq.~(\ref{SE}) exist, and that they are real, which is always true for the time-independent Schr{\"o}dinger equation in one dimension. 

Given a potential $V(x)$ we construct a new family of potentials $V_a(x)$, such that $V_a(x)=V(x)$ for $-a<x<a$, and $V(x)=\infty$ for $|x|>a$. For $a=\epsilon$ sufficiently small we have an infinite potential well and the wave functions are well known: $\psi_k^{(\rm o)}(x;\epsilon) \propto \sin(k\pi x/\epsilon)$, $k=1,2,\ldots$, and $\psi_k^{(\rm e)}(x;\epsilon) \propto \cos((2k +1) \pi x / 2\epsilon)$, $k=0,\,1,\,2,\,\ldots$, where the superscripts (e) and (o) refer to the even/odd parity of the wavefunctions. Clearly, $\psi_n(x;\epsilon)$ has $n-1$ nodes between $-\epsilon$ and $\epsilon$ where it vanishes. We first focus on the ground state wave function, which we may assume to be positive, because there are no nodes, that is, $\psi'_1(-\epsilon; \epsilon) > 0$ and $\psi'_1(\epsilon;\epsilon) < 0$.\cite{prime} Imagine that we separate the ``walls'' by increasing $a$. The wavefunction $\psi_1(x;a)$ will become a better and better approximation to the true ground state wave function, starting from a wavefunction without any nodes. Can this wavefunction develop a node between $\pm a$, for some value of $a$? If so, there are two possibilities: either (1) at least one of the derivatives at $\pm a$ must change sign, or (2) the derivatives at $\pm a$ do not change sign, but the wavefunction develops two zeros through its decrease at some point between $-a$ and $a$. In both cases there will be a critical value of $a$ such that the wavefunction and its first derivative vanish at the same point. In case (1) the wavefunction and its derivative vanish at one of the boundary points, because the boundary condition implies that the wavefunction vanishes at the boundaries, and if the derivative changes sign, then it must be zero for the critical value of $a$. In case (2) there must be a value of $a$ such that the wavefunction touches the real axis just before the wavefunction dips down and develops two zeros, and therefore has zero value and zero slope there. Because the Schr{\"o}dinger equation is a linear, second-order, ordinary differential equation, it has a unique solution, given the value of the function and its derivative at the same point. But if the solution and its first derivative are zero at the same point, we conclude that the wavefunction must be identically zero, because $\psi = 0$ satisfies the differential equations and the vanishing conditions of the function and its derivative. Because we assume that there is always a nontrivial solution of the Schr{\"o}dinger equation for any value of $a$, we conclude that in both cases the ground state function can develop no nodes. The same reasoning shows that because the $n$th wavefunction starts with $n-1$ nodes, their number cannot increase nor decrease, because to develop a new zero or lose a zero, the wavefunction must go through one of the stages described in cases (1) and (2) as $a$ increases. 

There are cases where the potential can support only a finite number of bound states, say $N$. What happens to the wave functions for $n>N$? As we separate the walls, some of the wave functions go to zero because the separation between the zeros increases as $a \to \infty$, and the positive energy states become the continuum spectrum. As an example, consider the potential well with two walls, as discussed in Ref.~\onlinecite{Fl}. 

It is straightforward to show that between two consecutive zeros of $\psi_{n_1}$, there is a zero of $\psi_{n_2}$ for any $n_2 > n_1$. This statement can be proved as follows. From the Schr{\"o}dinger equation for $\psi_{n_1}$ and $\psi_{n_2}$, we see that 
\begin{equation}
(\psi_{n_1}' \psi_{n_2} - \psi_{n_2}' \psi_{n_1})'=(E_{n_2}-E_{n_1})\psi_{n_1}\psi_{n_2}. \label{eq2}
\end{equation}
If $\psi_{n_1}$ has two consecutive zeros at $x_1$ and $x_2$, we may assume that $\psi_{n_1}$ is positive in the interval between the two zeros. If $\psi_{n_2}$ does not vanish in this interval, we may also assume it is positive. If we ntegrate Eq.~(\ref{eq2}) from $x_1$ to $x_2$ and note that that $\psi'_{n_1}(x_1)>0$ and $\psi_{n_1}'(x_2)<0$, we obtain
\begin{equation}
\psi_{n_1}'(x_2) \psi_{n_2}(x_2)-\psi_{n_1}'(x_1) \psi_{n_2}(x_1)=(E_{n_2}-E_{n_1})\!\int_{x_1}^{x_2}\!dx\,\psi_{n_1}(x)\psi_{n_2}(x).
\end{equation}
Because the left-hand side is negative definite and the right-hand side is positive definite, we arrive at a contradiction. Therefore $\psi_{n_2}$ has a zero between two consecutive zeros of $\psi_{n_1}$.

The method presented here can be used to establish that the ground state wavefunction has no nodes in dimensions higher than one. All that is necessary is to consider an infinite potential well around the origin and separate the walls. Because the ground state wavefunction starts with no nodes, it can develop no nodal lines, for reasons similar to those we have discussed.

\begin{acknowledgements}
I would like to thank L.\ Moriconi and C.\ Tomei for useful discussions. The remarks of two anonymous referees are gratefully acknowledged, especially for pointing out Ref.~\onlinecite{M} and correcting the wavefunctions for the infinite potential well. This work has been partially supported by Faperj.
\end{acknowledgements}


\begin{thebibliography}{99}
\bibitem{note} We will call the ground state function $\psi_1$ instead of the more common notation $\psi_0$.
\end{thebibliography}
\end{document}